\documentclass[%
reprint,
onecolumn,
superscriptaddress,
amsmath,amssymb,
aps,
11pt,
pre,
longbibliography
]{revtex4-1}

\usepackage{geometry}
\usepackage{tgtermes}

\usepackage{graphicx}
\usepackage{dcolumn}
\usepackage{bm}
\usepackage{color}
\usepackage[table]{xcolor}
\usepackage[utf8]{inputenc}

\begin{document}

\title{Chaotic and regular instantons in  helical shell models of turbulence\footnote{Version accepted for publication (postprint) on Phys. Rev. Fluids 2, 034606 – Published 29 March 2017}}

\author{Massimo De Pietro}
\affiliation{Dipartimento di Fisica and INFN,
  Universit\`a ``Tor Vergata", Via della Ricerca Scientifica 1,
  00133 Roma, Italy.}  
  
\author{Alexei A. Mailybaev}
\affiliation{Instituto Nacional de Matem\'atica Pura e Aplicada, Rio de 
Janeiro, Brazil}

\author{Luca Biferale}
\affiliation{Dipartimento di Fisica and INFN,
	Universit\`a ``Tor Vergata", Via della Ricerca Scientifica 1,
	00133 Roma, Italy.}  

\date{\today}

\begin{abstract}
Shell models of turbulence have a finite-time blowup in the inviscid limit, i.e., the enstrophy  
diverges while the single--shell velocities stay finite.  
The signature of this blowup is represented by self-similar instantonic 
structures traveling coherently through the inertial range. 
These  solutions might influence the energy transfer and the anomalous 
scaling properties empirically observed for the forced and viscous models. 
In this paper we present a study of the instantonic solutions for 
a set of four shell models of turbulence based on the exact decomposition of the 
Navier-Stokes equations in helical eigenstates. We find that depending on the 
helical structure of each model, instantons are chaotic or regular. 
Some instantonic solutions tend to recover mirror symmetry for scales small 
enough. Models that have  anomalous scaling develop regular 
non chaotic instantons. Conversely, models that have non anomalous 
scaling in the stationary regime are those that have chaotic 
instantons. The direction of the energy carried by each single 
instanton tends to coincide with the direction of the energy cascade in the 
stationary regime. Finally, we find that whenever the small-scale stationary statistics is intermittent, the instanton is less steep than the dimensional Kolmogorov scaling, independently of whether or not it is chaotic. Our findings further 
support the idea that instantons might be crucial to describe some aspects of the multi-scale anomalous statistics of shell models.

\end{abstract}

\maketitle

\section{Introduction}
The phenomenological 
Kolmogorov theory is able to catch the basic constituents of the energy 
transfer mechanisms in homogeneous and isotropic turbulence, but it falls 
short in explaining subtle effects such as intermittency, i.e., the existence of 
anomalous scaling laws for velocity increments  in the limit of high Reynolds 
numbers ~\cite{frish_turbulence}, where the Reynolds number $\text{Re}$ measures 
the relative importance of inertial and viscous effects. When $\text{Re} \rightarrow 
\infty$,  a wide separation opens between the scale where forcing and viscous 
mechanism act, making the problem computationally hard and analytically 
intractable.
Up to now, all attempts to attack the problem for the three--dimensional Navier-Stokes equation (NSE) have failed. 
As a result, many approximate approaches have been developed in 
order to gain insights into the transfer mechanisms in turbulent flows. A 
successful approach is 
represented by shell models \cite{obukhov1971some, gledzer1973system, 
desnianskii1974evolution, yamada1988inertial, jensen1991intermittency, 
Lvov_1998_improved_shellmodels, biferale2003shell, bohr2005dynamical, 
ditlevsen2010turbulence}, especially concerning the existence of intermittency and anomalous scaling laws.

Shell models of turbulence are dynamical models that mimic the NSE in the wave-number space. They are based on a  strong 
reduction of the number of degrees of freedom, dividing and discretizing the 
Fourier space into a number of shells equally spaced on a logarithmic scale 
$k_n = k_0 \lambda^n$ (a common choice is $\lambda=2$ and $k_0=1$). Only a few 
representative variables $u_n$ are kept for each shell of wavenumbers. Each variable is meant to 
represent
a typical velocity fluctuation of the original three--dimensional Navier-Stokes field 
$\delta_r v$  at scale $r \sim 1/k_n$. 
In this way, a large separation of scales can be achieved with relatively few 
variables. Furthermore, inspired by the Kolmogorov phenomenology for the 
direct energy transfer, these models consider only local interactions in 
Fourier space, connecting dynamical evolution between three generic neighboring 
modes  $k_n, k_{n+1}, k_{n+2}$. Finally, the models are built in such a way as
to have the same inviscid invariants of the original NSE: 
energy and helicity for models of three--dimensional (3D) turbulence or energy and enstrophy for 
2D turbulence.

The success of these models lies in the fact that despite the huge 
simplifications, they still share many properties with the original 
Navier-Stokes turbulence, including the development of anomalous scaling laws 
with values of the scaling exponents very close to the ones measured in 3D 
turbulence \cite{jensen1991intermittency, pisarenko1993further, 
Lvov_1998_improved_shellmodels, biferale2003shell}. Though the 
turbulence even in such simplified models is far from being understood, shell models 
remain important in fluid dynamics for accessing detailed properties of energy 
transfer mechanisms.

In particular, in \cite{mailybaev2013blowup} the issue of intermittency 
 was studied in one popular shell model~\cite{Lvov_1998_improved_shellmodels} and it was argued that anomalous scaling 
exponents of velocity moments can be related to the scaling and statistics of instantons. 
Instantons are particular solutions of the inviscid equations of motion, 
intimately connected to the finite time blowup of the model with an infinite number of shells~\cite{dombre1998intermittency, daumont2000instanton}. 
In the turbulent velocity field they are represented by coherent structures 
that traverse the inertial range towards large wave numbers. 
 In this work, we attribute the word instanton to a self-similar 
inviscid structure localized in both time and scale, which is different from  the 
viscous instantonic solutions generated within the Martin-Siggia-Rose 
formalism and widely studied  for the original three-dimensional NSE 
and for Burgers equations \cite{Falkovich1996, Gurarie1996, 
	Balkovsky1997, dombre1998intermittency, Biferale1999, daumont2000instanton, 
	Chernykh2001, l2002quasisolitons, mailybaev2013blowup, Grafke2015}. 

In the following we study the structure of instantonic solutions in four different classes of shell models~\cite{benzi_1996_Helical_shell_models} generalized
to have a closer analogy with the original structure of the NSE decomposed on a helical Fourier basis \cite{waleffe_1992_The_nature_of_triad_interactions}. 
Helical decomposition of the NSE is 
useful to disentangle triad interactions that preferentially transfer energy 
to small or to large scales (forward and backward energy cascades).   This 
statement was 
recently supported in direct numerical simulation of the NSE with appropriate dynamical mode 
reduction \cite{biferale2012inverse, Biferale2013, sahoo2015disentangling, 
alexakis2016helically} and in the equivalent helical version of shell models 
\cite{depietro2015inverse}. Let us note that instantonic solutions were shown to be closely related to the events preceding a shock formation in compressible flows  \cite{mailybaev2012renormalization}, 
justifying their relevance also  for 
realistic hydrodynamical systems in the continuum. Such a relation for incompressible flows, as well as for the case of a chaotic instanton, is unknown. Chaotic instantons may turn out to be  useful for describing an non regular behavior in 3D Euler equations, in relation to the open problem of finite-time blowup, and they are also conjectured to describe the Belinsky-Khalatnikov-Lifshitz singularity solution for Einstein’s field equations of gravitation \cite{belinskii1970oscillatory}. 

The paper is structured as follows. In Secs. 
\ref{sec:theor_intro_helical_sm} and \ref{sec:theor_intro_blowup} we  
review the general concepts of helical shell models and define the 
instantonic solutions for such models. In Secs. 
\ref{sec:results_dynamics_and_spectra}, \ref{sec:results_helicity}, and 
\ref{sec:results_etransfers} we  show results from numerical simulations, 
concerning different aspects: the general dynamics of the instantons, their 
helical structure, and the energy transfers they induce, respectively. Finally, in Sec. 
\ref{sec:conclusions} we  discuss our findings and summarize the 
connections between the instantonic solutions and the stationary dynamics of 
shell models and real turbulence. 

\section{Helical shell models}
\label{sec:theor_intro_helical_sm}
The three-dimensional incompressible Navier-Stokes equations can be exactly decomposed on a base of 
positive and negative polarized helical waves 
\cite{waleffe_1992_The_nature_of_triad_interactions}.  
In Fourier space, this helical decomposition for a velocity field reads
\begin{equation}
\label{eq:waleffe_ns_u_decomposition}
\mathbf{u}(\mathbf{k}) = u_{\mathbf{k}}^+ \mathbf{h}_{\mathbf{k}}^+ + 
u_{\mathbf{k}}^- \mathbf{h}_{\mathbf{k}}^- \, ,
\end{equation}
where for each wave vector $\mathbf{k}$, $\mathbf{h}_{\mathbf{k}}^+$, and 
$\mathbf{h}_{\mathbf{k}}^-$ are eigenvectors of the curl operator,
\begin{equation}
i \mathbf{k} \times \mathbf{h}_{\mathbf{k}}^s = s k \mathbf{h}_{\mathbf{k}}^s \,.
\end{equation}
Such vectors carry, respectively, positive and negative helicity and can 
be taken as
\begin{equation}
\label{eq:waleffe_h_s}
\mathbf{h}_{\mathbf{k}}^s = \boldsymbol{\nu}_{\mathbf{k}} \times 
\boldsymbol{\kappa} + s i \boldsymbol{\nu}_{\mathbf{k}} \, ,
\end{equation}
where $\mathbf{k} = k \boldsymbol{\kappa}$ and $\boldsymbol{\nu}_{\mathbf{k}}$ is an arbitrary vector orthogonal to $\mathbf{k}$. Then the two 
fields $u^+_{\mathbf{k}}$ and $u^-_{\mathbf{k}}$ are the projections on the 
$\mathbf{h}_{\mathbf{k}}^+$ and $\mathbf{h}_{\mathbf{k}}^-$ directions of the 
Fourier coefficients of the velocity field.
Plugging decomposition (\ref{eq:waleffe_ns_u_decomposition}) into the 
non linear term of the NSEs, one can distinguish eight possible 
non linear triadic interactions depending on the signs of the corresponding 
helical projections \cite{waleffe_1992_The_nature_of_triad_interactions}. 
Four out of eight interactions are independent, because the interactions with 
reversed helicities are identical. The four structures of interacting triads 
will be labeled SM1-4 and they are summarized in the second column of Table \ref{tab:helical_interactions}. 

It is possible to construct four different  shell models with  a helicity structure
analogous to that of the  four sub-classes of the original  NSEs \cite{benzi_1996_Helical_shell_models, depietro2015inverse}:
\begin{align}
\label{eq:sabra_helical_general_up}
\dot{u}_n^+ = & \, i (a k_{n+1} u_{n+2}^{s_1} u_{n+1}^{s_2*} + b k_{n} 
u_{n+1}^{s_3} u_{n-1}^{s_4*} + c k_{n-1} u_{n-1}^{s_5} u_{n-2}^{s_6}) + f_n^+ - \nu k_n^2  u_n^+  \, ,\\ 
\label{eq:sabra_helical_general_um}
\dot{u}_n^- = & \, i (a k_{n+1} u_{n+2}^{-s_1} u_{n+1}^{-s_2*} + b k_{n} 
u_{n+1}^{-s_3} u_{n-1}^{-s_4*} + c k_{n-1} u_{n-1}^{-s_5} u_{n-2}^{-s_6}) + f_n^-  - \nu k_n^2   u_n^-   \, 
, 
\end{align}
where $n = 1,2,\ldots$ are shell indices and $u_n^+$ and $u_n^-$ are complex 
shell variables (speeds) corresponding to positive and negative helicity 
modes. The  helical indices $s_i = \pm$ and the coefficients $a,b,c$ can be 
found in Table \ref{tab:helical_interactions}. Note that model 
SM1 can be split into two identical fully uncoupled models for the 
variables $u_1^+,u_2^-,u_3^+,\ldots$ and $u_1^-,u_2^+,u_3^-,\ldots$. The same is true for model SM4, 
where the uncoupled models are  $u_1^+,u_2^+,u_3^+,\ldots$ and 
$u_1^-,u_2^-,u_3^-,\ldots$; models SM2 and SM3, on the 
contrary, cannot be decoupled.
\begin{table*}
	\caption{Structure and coefficients of the four helical models 
		\eqref{eq:sabra_helical_general_up} and 
		\eqref{eq:sabra_helical_general_um}. The second column lists classes of 
		helical interactions. Without loss of generality, we always choose 
		$a=1$. These $a$, $b$, and $c$ coefficients ensure energy and helicity 
		conservation.}
	\label{tab:helical_interactions} 
	\begin{tabular*}{\linewidth}{@{\extracolsep{\fill} } c | c c c c c c c c c}
		\toprule
		Model & Helical modes coupling & $s_1$ & $s_2$ & $s_3$ & $s_4$ & 
		$s_5$ & $s_6$ & $b$ & $c$ \\ \colrule
		SM1  &  ($u_{n}^+$, $u_{n+1}^-$, $u_{n+2}^+$) or ($u_{n}^-$, 
		$u_{n+1}^+$, $u_{n+2}^-$) &$+$&$-$&$-$&$-$&$-$&$+$&$-1/2$&$1/2$\\
		SM2  &  ($u_{n}^+$, $u_{n+1}^-$, $u_{n+2}^-$) or ($u_{n}^-$, 
		$u_{n+1}^+$, $u_{n+2}^+$) &$-$&$-$&$+$&$-$&$+$&$-$&$-5/2$&$-3/2$\\
		SM3  &  ($u_{n}^+$, $u_{n+1}^+$, $u_{n+2}^-$) or ($u_{n}^-$, 
		$u_{n+1}^-$, $u_{n+2}^+$) &$-$&$+$&$-$&$+$&$-$&$-$&$-5/6$&$1/6$\\
		SM4  &  ($u_{n}^+$, $u_{n+1}^+$, $u_{n+2}^+$) or ($u_{n}^-$, 
		$u_{n+1}^-$, $u_{n+2}^-$) &$+$&$+$&$+$&$+$&$+$&$+$&$-3/2$&$-1/2$\\
		\botrule
	\end{tabular*}
\end{table*}
In shell models 
(\ref{eq:sabra_helical_general_up}) and (\ref{eq:sabra_helical_general_um}) both 
the total energy $E$  and the total helicity $H$ are conserved for zero 
viscosity and zero forcing (just as in NSEs):
\begin{equation}
\label{eq:energy_def}
E = \sum_{n=1}^{\infty} E_n  \, , \quad 
H = \sum_{n=1}^{\infty} H_n  \, ,
\end{equation}
where the energy and helicity spectra are
\begin{equation}
\label{eq:energy_def_flux}
E_n = |u_n^+|^2 + |u_n^-|^2,\quad
H_n = k_n( |u_n^+|^2 - |u_n^-|^2 ).
\end{equation}

Note that any linear combination of models SM1-SM4 conserves 
the total energy and helicity. The coupling among the four models can be  
explicitly calculated so as to be consistent with the structure of the NSE  
\cite{rathmann2016role}.

\section{Finite-time blowup in the inviscid model}
\label{sec:theor_intro_blowup}
In this paper we are interested in understanding the propagation of 
fluctuations in the inertial range of scales, i.e., in the inviscid limit.
In such a limit, solutions of shell models are 
characterized by a finite-time infinite growth (blowup) of the 
enstrophy~\cite{dombre1998intermittency, constantin2007regularity, cheskidov2008blow, mailybaev2012renormalization}:
\begin{equation}
\Omega(t) \rightarrow \infty \quad \text{as} \, t \rightarrow t_c^-.
\end{equation}
For helical models, the enstrophy is defined as
\begin{equation}
\label{eq:enstrophy_def}
\Omega = \omega^2 = \sum_{n=1}^{\infty} k_n^2 ( |u_n^+|^2 + |u_n^-|^2 ),
\end{equation}
where we also introduced $\omega$ as the square root of the enstrophy for further convenience.
The dynamical signature of this blowup is a coherent structure that travels 
from small to large wave numbers in a self-similar manner. 

Furthermore, it is 
possible to find a suitable change of variables that turns the blowup 
solution into a steady-state traveling wave, which is much easier to study. Concerning the helical models, 
we introduce the renormalized time $\tau$ and shell variables $w_n^\pm$, respectively,
\begin{equation}
\label{eq:w_def}
\frac{d\tau}{dt} = \frac{\omega(t)}{\omega_0},\quad
w_n^\pm(\tau) = -\frac{i k_n u_n^\pm(t)}{\omega(t)/\omega_0},
\end{equation}
where $\Omega_0 = \omega_0^2$ is the initial enstrophy value at $t = \tau = 0$. 
These variables are designed such that their norm
\begin{equation}
\label{norm_ren}
\|w\|^2 = \sum_{n=1}^\infty \left(|w_n^{+}|^2+|w_n^{-}|^2\right) = \Omega_0 
\, ,
\end{equation}
is conserved.
With definitions (\ref{eq:w_def}) it is possible to rewrite the inviscid and 
unforced equations 
(\ref{eq:sabra_helical_general_up}) and (\ref{eq:sabra_helical_general_um}) in 
the form 
\begin{align}
\label{eq:sabra_helical_general_wp}
\left( \frac{d}{d \tau} + A(\tau) \right) w_n^+ = & \, a \lambda^{-2} 
w_{n+2}^{s_1} w_{n+1}^{s_2*} + b w_{n+1}^{s_3} w_{n-1}^{s_4*} - c \lambda^2 w_{n-1}^{s_5} w_{n-2}^{s_6} \, ,\\ 
\label{eq:sabra_helical_general_wm}
\left( \frac{d}{d \tau} + A(\tau) \right) w_n^- = & \, a \lambda^{-2} 
w_{n+2}^{-s_1} w_{n+1}^{-s_2*} + b w_{n+1}^{-s_3} w_{n-1}^{-s_4*} - c \lambda^2 w_{n-1}^{-s_5} w_{n-2}^{-s_6},
\end{align}
where 
\begin{equation}
A = \frac{1}{\omega} \frac{d \omega}{d \tau} \, .
\label{eq:A_def}
\end{equation}
For $\omega$ (square root of the enstrophy), we get
\begin{equation}
\omega(\tau) = \omega_0 \exp \left(\int_0^{\tau} A(\tau')d\tau'\right).
\label{A_to_Omega}
\end{equation}
Here we wrote $\omega$ as a function of $\tau$, which in turn is a function 
of original time $t$. Differentiating (\ref{norm_ren}) with respect to $\tau$ 
and using 
(\ref{eq:sabra_helical_general_wp}) and (\ref{eq:sabra_helical_general_wm}), one 
can get an explicit expression for 
\begin{equation}
A = \frac{1}{\Omega_0} \sum_n \mathrm{Re}\left(w_n^{+*} NLT^+_{n} + 
w_n^{-*} NLT^-_{n}\right) \, ,
\label{eq:A_def_2}
\end{equation}
where $NLT^+_{n}$ and $NLT^-_{n}$ represent the right-hand sides of Eqs. 
(\ref{eq:sabra_helical_general_wp}) and (\ref{eq:sabra_helical_general_wm}).

This renormalization completely removes the stiffness (exponential decrease 
in local timescale at increasing shell numbers $n$) of the original system, 
and maps the blowup limit $t \to t_c^-$ to the infinite limit $\tau \to 
\infty$, so the solutions are well-defined globally in the renormalized 
time $\tau$. Note that there is a one-to-one exact correspondence between 
solutions of the original and renormalized systems, for  $t < t_c$.

The blowup can be described asymptotically as an attractor of the 
renormalized dynamics~\cite{mailybaev2013bifurcations}. 
For instance, as the norm $\|w\|^2 = \Omega_0$ is conserved, the renormalized 
system 
(\ref{eq:sabra_helical_general_wp}) and (\ref{eq:sabra_helical_general_wm}) may 
have a solitary wave solution  
\begin{equation}
w_n^\pm = W^\pm(n-s\tau) \, ,
\label{eq:traveling_wave}
\end{equation}
where $s$ represents the wave speed and $W^\pm(\xi)$ are functions vanishing 
as $\xi \rightarrow \pm \infty$. 
Let us introduce the scaling exponent 
\begin{equation}
y = \log_\lambda \frac{\omega(\tau_1)}{\omega_0} > 0,\quad \tau_1 = 1/s,
\label{blowupcond}
\end{equation}
where the value $\tau_1$ is defined as the renormalized time in which 
solution (\ref{eq:traveling_wave}) travels over a single shell $n \mapsto 
n+1$. If $y > 0$,
then the traveling wave (\ref{eq:traveling_wave}) represents the self-similar 
finite-time blowup for the original shell variables $u_n^\pm$ given by 
(\ref{eq:w_def}) as~\cite{dombre1998intermittency, mailybaev2013bifurcations}:
\begin{equation}
u_n^\pm = i k_n^{y-1} U^\pm[k_n^y(t-t_c)] \, ,
\label{eq:u_instanton}
\end{equation}
where
\begin{equation}
U^\pm(t-t_c) = \frac{\omega(\tau)}{\omega_0} \,W^\pm(-s\tau) ,
\label{slefsimilar}
\end{equation}
\begin{equation}
t_c = \int_0^\infty \exp \frac{\omega(\tau)}{\omega_0} \,d\tau' < \infty .
\label{blowuptime}
\end{equation} 
Here the condition $y > 0$ is necessary to ensure the convergence of the 
integral (\ref{blowuptime}), i.e. the finiteness of the blowup time $t_c$.

\section{Regular and chaotic instantons}
\label{sec:results_dynamics_and_spectra}
	
We have performed a series of numerical integrations of Eqs.  
(\ref{eq:sabra_helical_general_wp}) and (\ref{eq:sabra_helical_general_wm}), 
using a standard fourth-order Runge-Kutta scheme. For each model we made a 
number of simulations with different initial conditions. In the initial 
conditions, energy was distributed uniformly over a small interval of 
shellnumbers $n = 10,\ldots,14$; the velocity was zero elsewhere. For every 
initial condition, the energy $E = 1$ and helicity $H = 1.55$ were the same, 
while the phases of velocity variables were random. Since the 
stiffness characterizing the original shell model equations is removed in the 
renormalized description, we were able to study a very large range of shell 
numbers ($N=120$ total shells are used in most simulations) with a 
shell-to-shell ratio $\lambda=2$. Each simulation was stopped as soon as the 
energy reached the highest wave number. 
Given the possibility to achieve extremely high 
wave numbers, care must be taken when measuring the helicity $H$ or other 
helicity-sensitive quantities [in general, all observables of the form 
$k_n^\alpha(|u_n^+|^\beta - |u_n^-|^\beta)$], because huge cancellations 
might take place at high wave numbers and quadruple precision arithmetic is 
 required for large $N$.

\begin{figure}
\includegraphics[width=0.75\linewidth]{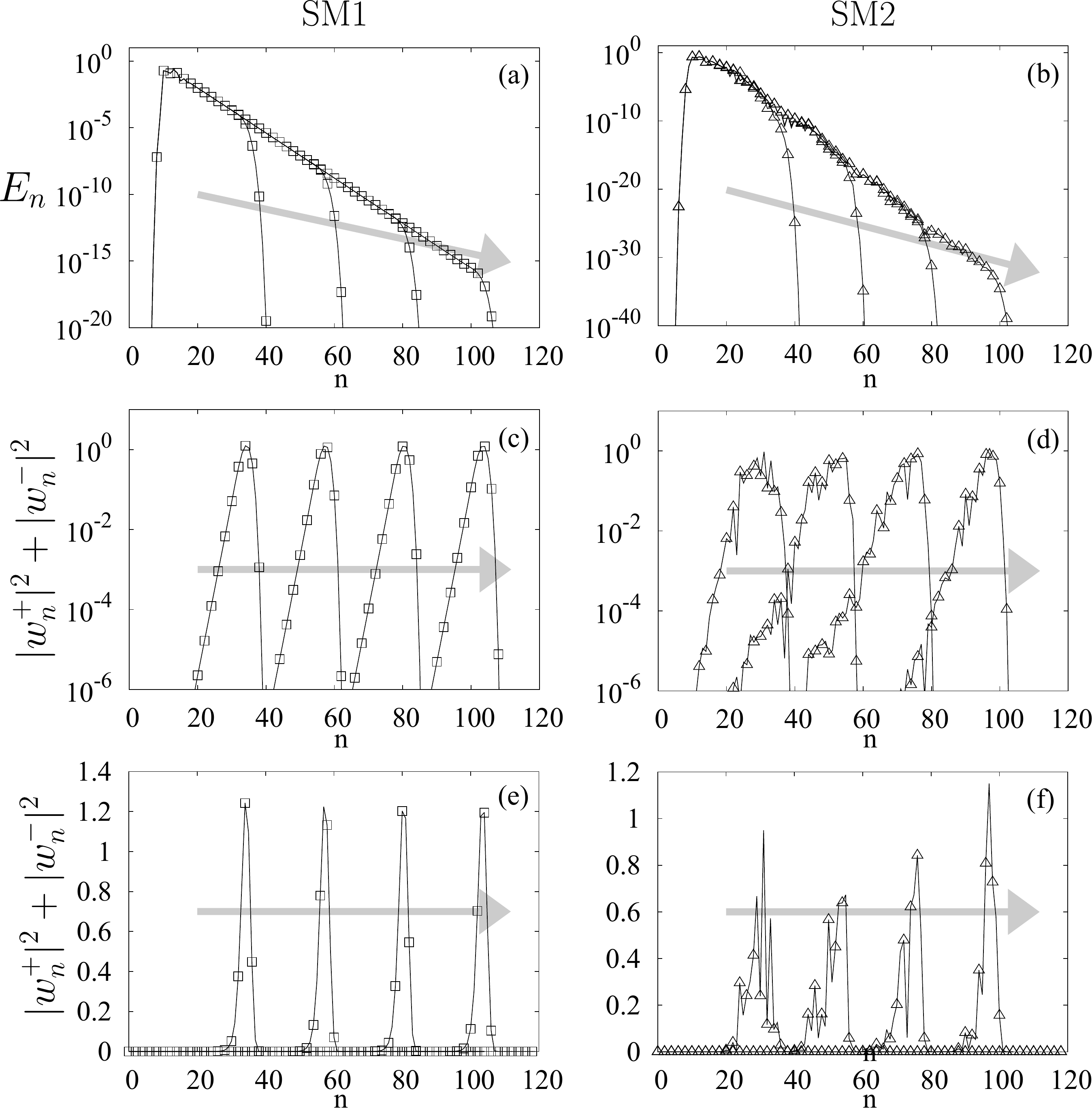}
\caption{Snapshots of the solution of Eqs. 
(\ref{eq:sabra_helical_general_wp}) and (\ref{eq:sabra_helical_general_wm}) at different, equally separated, moments in the renormalized time 
$\tau$, for two models with a regular and a chaotic instanton: energy 
spectrum for (a) model SM1 and (b) model SM2 and norm 
spectrum $|w_n^+|^2+|w_n^-|^2$ for (c) model SM1 and (d) 
model SM2. Panels (e) and (f) show the same curves as (c) and (d), 
without logarithmic scale on the $y$ axis. The arrows in the background show 
the direction of increasing $\tau$.}
\label{fig:1}
\end{figure} 
	
Two types of limiting behavior were observed at large $\tau$, depending on 
the model. Models SM1 and SM3 exhibit an attractor in the 
form of a traveling wave, which moves toward larger shell numbers $n$ keeping 
a constant shape $W^\pm(\xi)$ and speed $s$ [see 
Eq.~(\ref{eq:traveling_wave})]. Models SM2 and SM4 on the 
other hand, show chaotic behavior with a  solution moving in the same 
direction of large $n$. 
Figure \ref{fig:1} shows two representative cases of regular 
(left) and chaotic (right) dynamics. 
Both the energy $E_n$ and the norm $|w_n^+|^2+|w_n^-|^2$ spectra at each 
shell are shown at equally separated moments in renormalized time $\tau$, 
showing clearly the traveling wave nature of the solution. In the first case, 
the wave has a constant profile and we say that the instanton is regular, 
while in the second case the profile fluctuates chaotically and we call the 
instanton chaotic.

At each time, the dynamics is effectively confined to a finite number of 
shells in the front of the propagating pulse, while in the tail of 
the solution, i.e., at smaller shell numbers $n$, the dynamics is frozen 
due to much larger characteristic time scales [see 
Figs.~\ref{fig:1}(a) and~\ref{fig:1}(b)]. Thinking in terms of the original time 
$t$, the dynamics is localized in the instants immediately preceding the 
blowup time $t_c$. 

\begin{figure}
\includegraphics[width=0.75\linewidth]{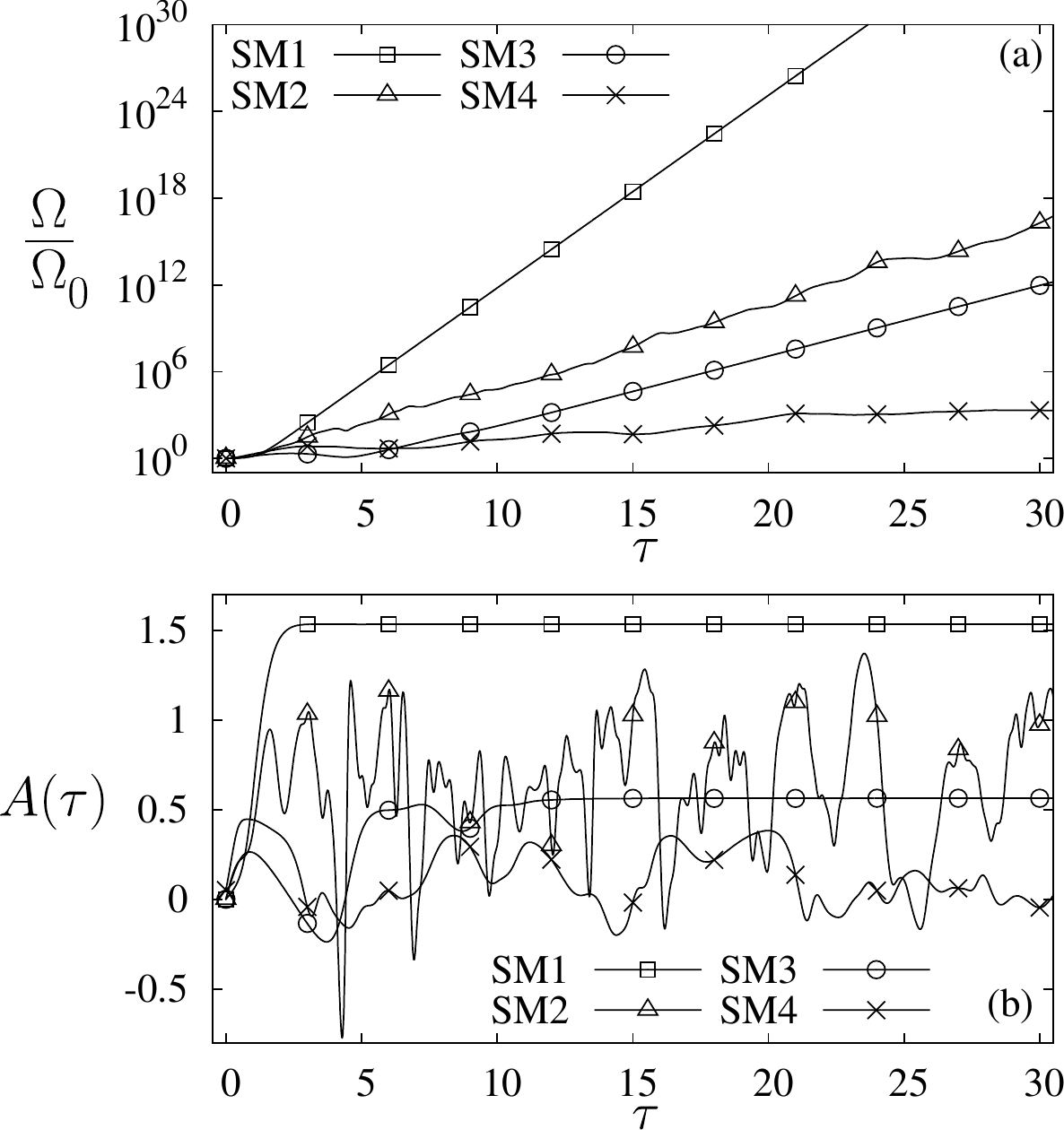}\\
\caption{(a) Relative enstrophy $\Omega/\Omega_0$ growth with 
renormalized time $\tau$ for different models in a single realization 
of the instanton. (b) Logarithmic derivative $A = 
\frac{1}{\omega}\frac{d\omega}{d\tau}$ for different models. The 
curves stabilize near specific values (regular instantons, 
SM1 and SM3) or oscillate chaotically (chaotic 
instantons, SM2 and SM4). }
\label{fig:2} 
\end{figure}

Figure \ref{fig:2} shows the relative enstrophy growth 
$\Omega/\Omega_0$ with $\tau$ and the corresponding logarithmic derivative $A = 
\frac{1}{\omega}\frac{d\omega}{d\tau}$ [see Eq.~(\ref{eq:A_def})] for the 
different models. We clearly distinguish two different behaviors. The 
enstrophy growth is exponential on average for large $\tau$. However, the 
growth rate $A$ stabilizes near specific values for models SM1 and 
SM3, where the attractor in the renormalized system is a traveling 
wave. On the contrary, a chaotically pulsating $A$ is observed for models 
SM2 and SM4, where the attractor is chaotic.

\begin{figure}
\includegraphics[width=0.75\linewidth]{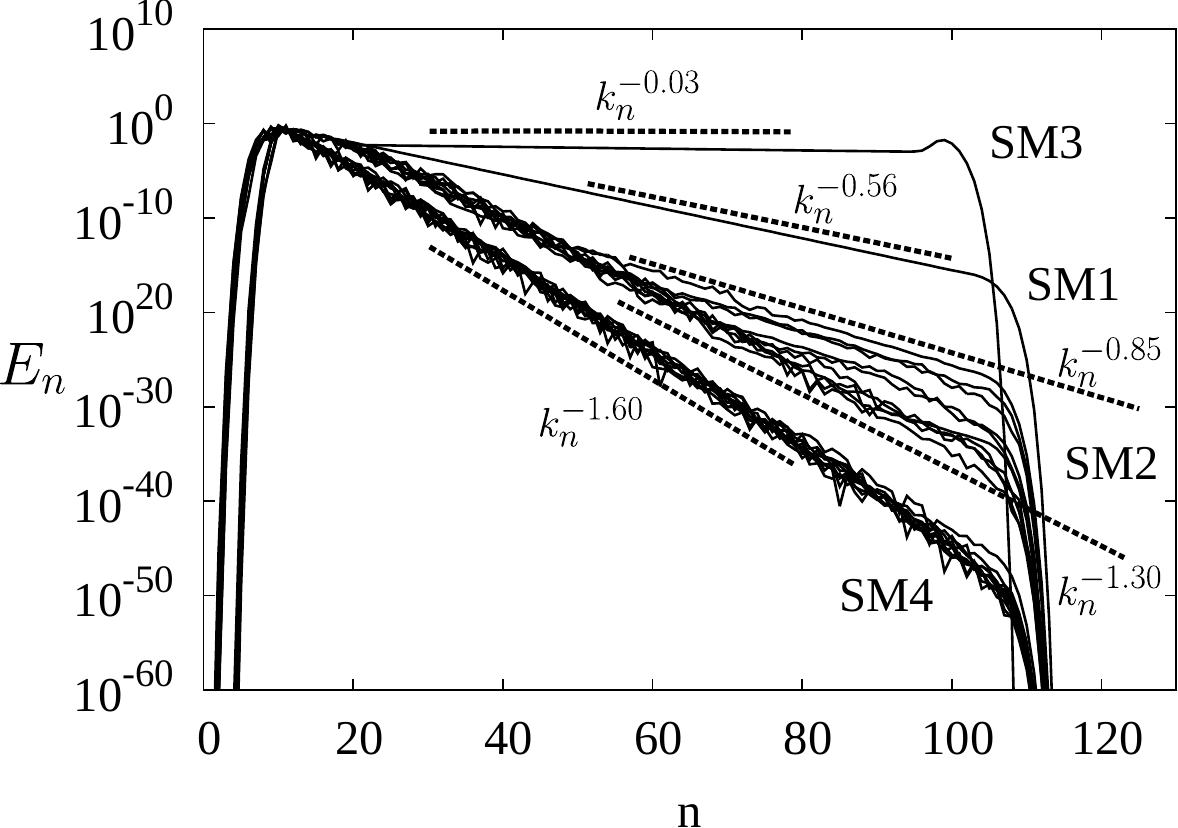}
\caption{Late-time ($t \rightarrow t_c^-$) energy spectrum $E_n$ of the 
instantons for different helical models. Multiple realizations are 
shown for the chaotic instantons (models SM2 and 
SM4). The dotted lines show scaling laws that best fit the 
different curves.}
\label{fig:3} 
\end{figure} 
	
\begin{figure}
\includegraphics[width=0.75\linewidth]{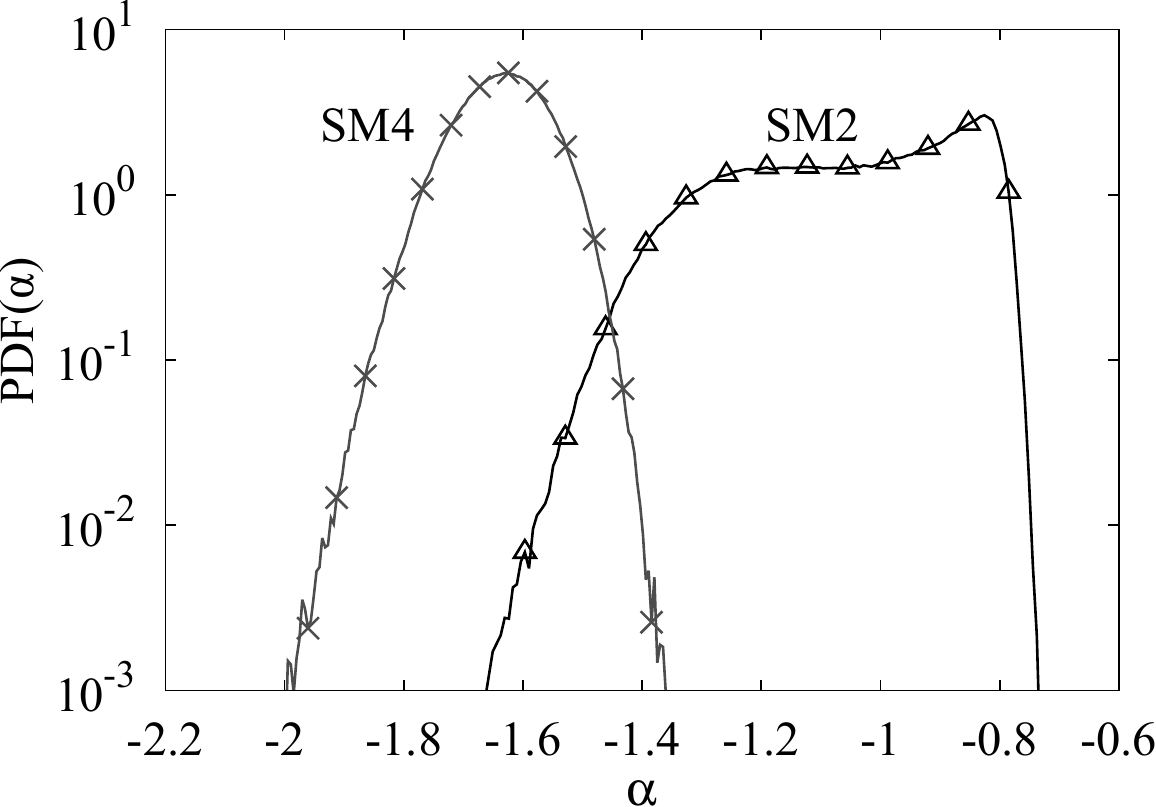}
\caption{PDF of local scaling exponents $\alpha$ of the energy spectrum $E_n$ 
(see Fig. \ref{fig:3}) for models SM2 and SM4.}
\label{fig:4} 
\end{figure}

By approaching the infinite shell number $n \to \infty$ as $\tau \to \infty$ 
(corresponding to $t \to t_c^-$), the energy gets a specific distribution 
over the whole range of scales, as shown in~Fig.~\ref{fig:3}. 
The regular instanton (models SM1 and SM3) leaves behind an 
asymptotically exact power-law energy spectrum, while the chaotic instanton (models 
SM2 and SM4) leads to the power-law energy spectrum only on 
average, and a fluctuating component remains at all scales. The scaling 
exponents of the energy spectra vary greatly from model to model, a hint 
that the different helical non linear interactions (models) may have a 
different degree of influence over the dynamics of the whole system when 
coupled together. Figure~\ref{fig:3} presents results for a 
number of different simulations, which have the same initial amplitudes for 
the $u_n^\pm$ limited to $n \in [10,14]$ but with different phases, randomly chosen.
 An interesting feature 
that distinguishes the chaotic instantons of model SM2 from the 
one of model SM4 is the increasing spread of the energy profiles at small scales shown by the former. Simulations with larger total shells $N$ indicate that this spread may be explained as the 
intermittency phenomenon, in the context of dynamical 
systems~\cite{ott2002chaos}: In the renormalized variables, the wave undergoes 
irregular jumps between periodic and chaotic dynamics, and among regimes 
characterized by different scalings. This is shown in Fig. 
\ref{fig:4}, where we plot the probability density function (PDF) of the  local scaling exponents 
$\alpha$ of the energy spectrum $E_n \sim k_n^\alpha$ for 
the two models SM2 and SM4. The scaling exponents $\alpha$ are related to 
$y$ in Eq. (\ref{eq:u_instanton}) by $\alpha = 2(y-1)$. They are calculated by performing 
a power-law fit on several sections, 40 shells long, taken from the energy 
spectrum curves (Fig. \ref{fig:3}) (limited to $20 \le n \le 100$). As one can see, while model SM4 
has a  distribution peaked around $\alpha = -1.6$, for model SM2 we 
have a wider spectrum of values with a strong peak around $\alpha=-0.85$ and a less pronounced peak around $\alpha = -1.3$.

\section{Helical structure}
\label{sec:results_helicity}
Let us now analyze the helical component of the instantonic solutions. 
Model SM1 being made of two decoupled Sabra models, will  develop 
also decoupled instantons for each submodel 
with  different blowup times that  depend on the initial condition.  As a 
result,  only the fastest instanton 
will dominate the dynamics asymptotically and the connection 
among helicity and 
energy  spectrum is trivial: $H_n = (-)^n k_n E_n$. The 
very same happens for model SM4, with the only 
difference that $H_n =  k_n E_n$ (or $H_n =  -k_n E_n$).

\begin{figure}
\includegraphics[width=0.75\linewidth]{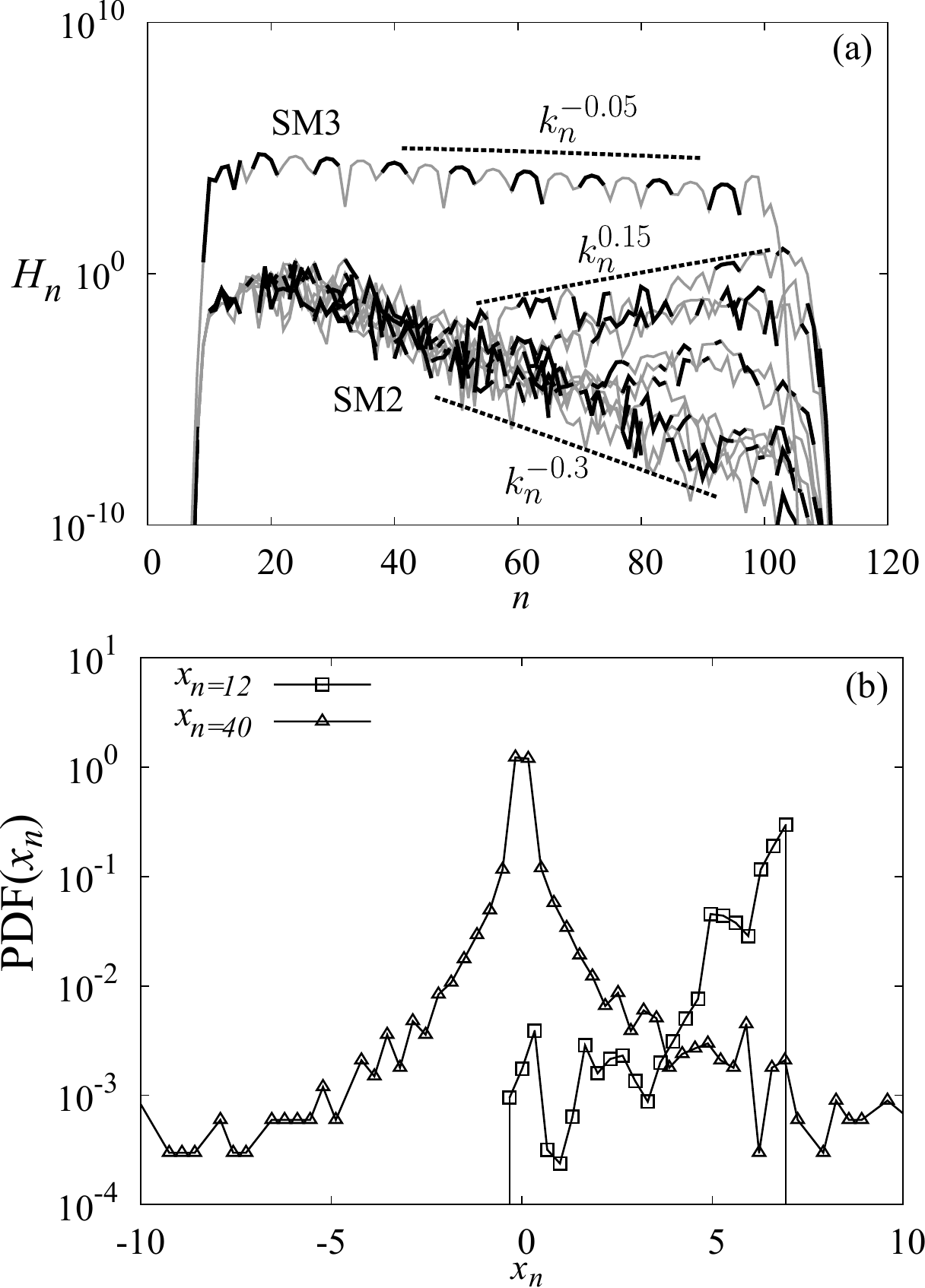}
\caption{(a) Late-time ($t \rightarrow t_c^-$) helicity spectra $H_n$ of the 
instantons for different helical shellmodels. Multiple realizations are shown 
for the chaotic instantons (model SM2). The two different models are 
shifted vertically for clarity The segments in gray represent negative 
values. (b) Late-time PDF of $x_n \equiv \frac{H_n}{\sqrt{\langle H_n^2 \rangle - \langle H_n \rangle^2}}$ for model SM2 at 
different shellnumbers $n$.}
\label{fig:5} 
\end{figure}

On the other hand, in models SM2 and SM3 all positive and 
negative helical modes are coupled, and the dynamics is richer.
 In Fig. \ref{fig:5}(a) 
we show the helicity 
spectrum for models SM2 and SM3 at a late time. We 
immediately notice that for model SM3 there is a fast recovery of 
parity invariance, $u_n^{\pm} \mapsto u_n^{\mp}$, as suggested by the alternation of positive (black) and negative (gray) signs in $H_n$.  A further 
confirmation of this recovery comes from the power-law scaling
\begin{equation}
	E_n = e_nk_n^{-\xi_E},\quad
	H_n = h_nk_n^{-\xi_H},
	\label{extr1}
\end{equation}
where $e_n$ and $h_n$ are $O(1)$  functions of $n$.
Exploiting (\ref{extr1}) in the helical 
decomposition (\ref{eq:energy_def_flux}), we can write 
 \begin{align}
	|u^+_n|^2 & = (e_n k_n^{-\xi_E} + h_n k_n^{-\xi_H-1})/2, \\
	|u^-_n|^2 & = (e_n k_n^{-\xi_E} - h_n k_n^{-\xi_H-1})/2,
\end{align}
with a power law for the relative helicity 
\begin{equation}
	\frac{|u^+_n|^2-|u^-_n|^2}{|u^+_n|^2+|u^-_n|^2}  \sim k_n^{\xi_E-\xi_H-1}.
	\label{extr2}
\end{equation}
Looking at Figs.~\ref{fig:3} and \ref{fig:5}(a), we conclude that 
model SM3 with $\xi_E \approx \xi_H$
has a strong recovery of mirror symmetry for small 
scales with the power law $k_n^{\xi_E-\xi_H-1} \approx
k_n^{-1}$. 

On the other hand, for model SM2 one has $\xi_E \approx \xi_H+1$. 
Hence, the chaotic behavior does not produce an exact cancellation of the 
leading mirror-symmetric terms and we 
observe $k_n^{\xi_E-\xi_H-1} \sim 1$ in (\ref{extr2}).
Nevertheless, the PDF of the helicity at 
different shell numbers indicates that even model SM2 eventually
recovers parity invariance. As shown in Fig. \ref{fig:5}(b), the PDF of $H_n$ is strongly skewed at 
shellnumbers where the (helical) initial condition is 
nonzero ($n=12$), while the same PDF becomes more and more symmetric at increasing 
$n$.
It is then argued that model SM2 will recover parity symmetry 
in a statistical sense when averaged over different instantonic solutions.
However, given the huge fluctuations in the energy and helicity spectra, 
this test would require an extremely high number of instantons to converge.

\section{Properties of instantons vs. developed turbulent dynamics}
\label{sec:results_etransfers}
In this section we study how the dynamics of the instantons compare with the 
stationary dynamics obtained in the same models with a forcing and viscous 
dissipation.
In particular, we are interested in understanding whether a 
correlation exists between the direction of the stationary energy transfer and
the transfer properties of instantonic solutions, together with issues 
connected to the anomalous scaling of the full stationary solutions.

In the stationary case, the energy transfer has been already studied in the 
literature \cite{ditlevsen1996cascades, benzi_1996_Helical_shell_models, 
depietro2015inverse}, showing that models SM1--SM3 have mainly 
a forward energy cascade (from large to small scales) while model SM4 have a backward energy 
transfer but close to a quasiequilibrium state \cite{ditlevsen1996cascades, 
gilbert2002inverse}. Recently, a model (SM2E) with the same helical structure of SM2 and 
second-neighbor interactions among modes  $k_n,k_{n+2},k_{n+3}$  was 
introduced in order to get a well developed inverse energy cascade regime, motivated by theoretical arguments based on the structure of the triadic interactions  \cite{depietro2015inverse}:
\begin{align}
\label{eq:sabra_helical_2p2_up}
\dot{u}_n^+ = & \, i (a k_{n+2} u_{n+3}^{-} u_{n+2}^{-*} + b k_{n} 
u_{n+1}^{+} u_{n-2}^{-*} + c k_{n-1} u_{n-1}^{+} u_{n-3}^{-}) + f_n^+ - \nu k_n^2  u_n^+  \, ,\\ 
\label{eq:sabra_helical_2p2_um}
\dot{u}_n^- = & \, i (a k_{n+2} u_{n+3}^{+} u_{n+2}^{+*} + b k_{n} 
u_{n+1}^{-} u_{n-2}^{+*} + c k_{n-1} u_{n-1}^{-} u_{n-3}^{+}) + f_n^- - \nu k_n^2  u_n^-   \, , 
\end{align}

where, for $\lambda=2$, the model constants are $a=1$, $b=-9/4$, and $c=-5/4$.

Model SM2E also develops a chaotic instanton with a slope for the energy spectrum around $\alpha=-1.4$ (discussed later). 

\subsection{Stationary dynamics}
A summary of the energy transfer direction for all the five models considered 
here is given in Table \ref{tab:summary_dynamics}. 
In the same table we also summarize what is known about the  scaling 
properties of the stationary dynamics for all models.
Scaling is here intended in terms of the structure functions for the 
full forced and viscous dynamics, defined as
\begin{equation}
\label{eq:sf-stationary}
S_p(k_n) = \langle |u_n^+|^p + |u_n^-|^p \rangle \sim k_n^{-\zeta_p},
\end{equation}
where with $\langle \cdots \rangle$ we mean the average over the statistically stationary ensemble and by $\zeta_p$ we denote the scaling exponents.

For 
the case of three-dimensional Navier-Stokes turbulence, it is 
empirically known that the equivalent of (\ref{eq:sf-stationary}) written for velocity increments in real space, $\langle (\delta_r v)^p \rangle \sim r^{\zeta_p}$, develops anomalous corrections: The scaling exponents do not follow a linear dimensional law $\zeta_p - (p/3) \zeta_3 \neq 0$.

\begin{figure}
	\includegraphics[width=0.75\linewidth]{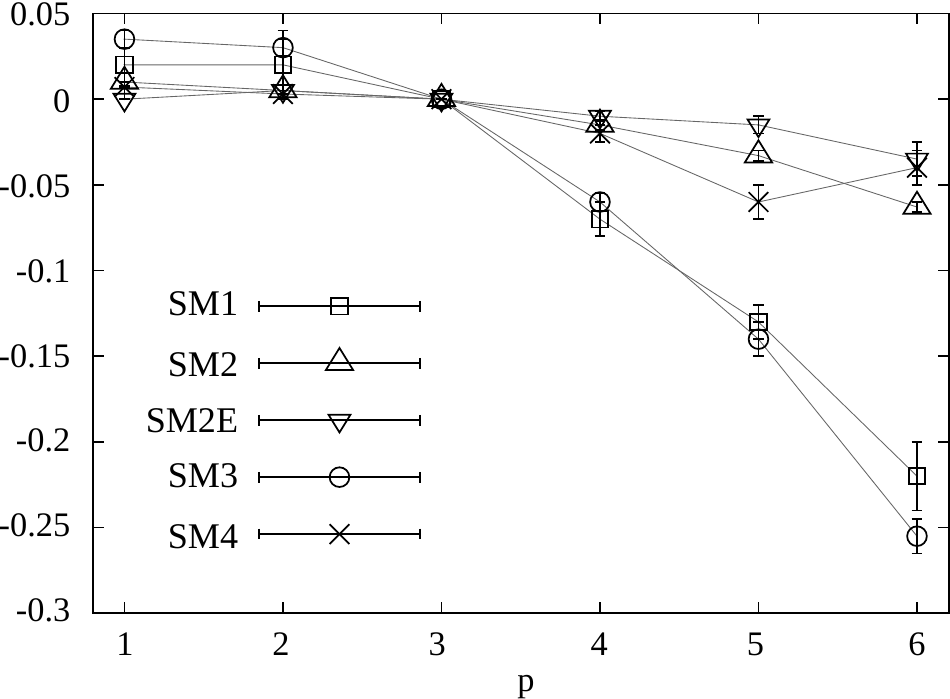}
	\caption{ Anomalous corrections $\Delta_p = \zeta_p - (p/3) \zeta_3$ 
			to the scaling exponents of the structure functions $S_p(k_n)$ 
			calculated in the forced-viscous regime, as functions of the order 
			$p$, for the various helical shell models.}
	\label{fig:6} 
\end{figure}

For the common choice $\lambda=2$ the shell models SM1 
and SM3
show anomalous exponents quantitatively very close to those of the full three-dimensional NSE.
Other models do not show intermittent behavior: 
SM2 has a non-intermittent forward cascade and the structure 
functions scale with exponents very close to ${\zeta_p} = p/3$; 
SM4 has a forward 
helicity cascade, as for the case of the NSE restricted to evolve only on 
a given sign of helical modes \cite{biferale2012inverse}, and the scaling 
exponents are very close to ${\zeta_p} = 2p/3$; finally 
SM2E has neither a forward 
energy cascade nor a forward helicity cascade, but the scaling exponents are 
still linear in $p$ (model SM2E actually shows a forward cascade of a third positive-definite invariant; see \cite{rathmann2016pseudo} for details). In Fig. \ref{fig:6} we summarize the anomalous corrections
	for the stationary structure 
functions in the presence of viscous and forcing terms for all models.
 
From the above considerations we notice that there exists a 
correlation between the presence of chaotic instantons and the absence of 
small-scale statistically stationary anomalous scaling, at least for the evolution of each helical shell model separately. Furthermore, we note that the absence of anomalous scaling for the stationary statistics is also correlated to the existence of instantons with an energy spectrum steeper than the dimensional Kolmogorov scaling $E_n \sim k_n^{-2/3}$. This follows from the condition $|\alpha| > 2/3$ in Table \ref{tab:summary_dynamics}.

\begin{figure}
	\includegraphics[width=0.75\linewidth]{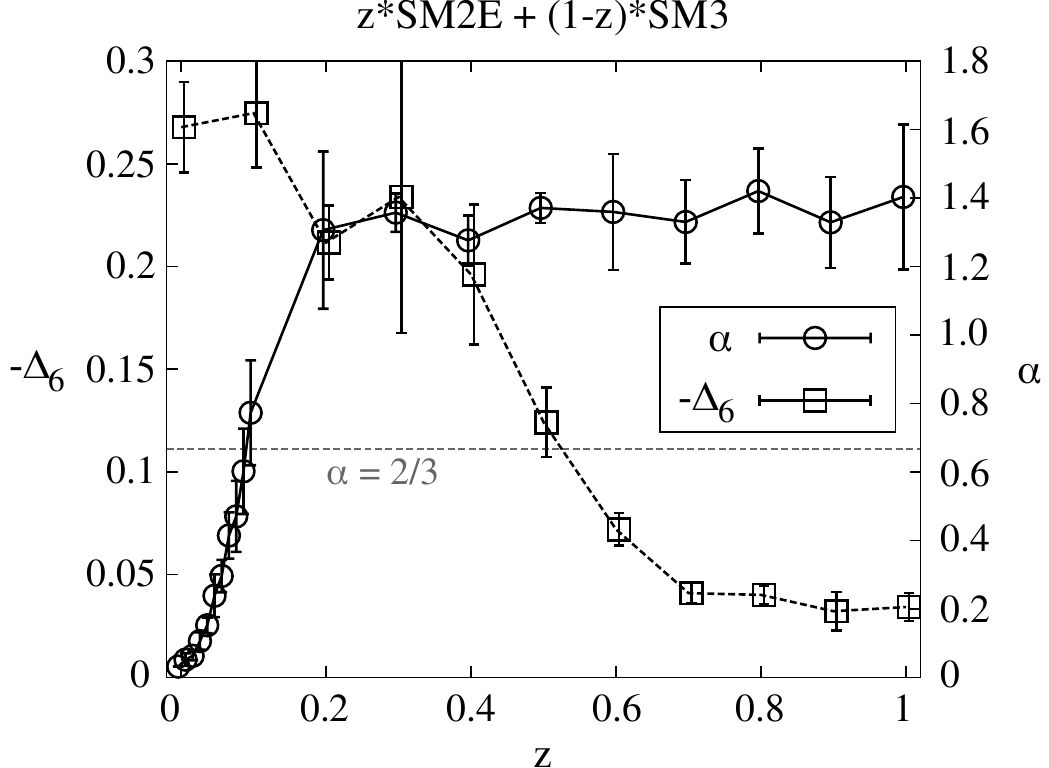}
	\caption{Comparison between the instantonic solution and the 
	forced-viscous system solution, for a linear combination of models 
	SM2E and SM3, with coupling coefficients $z$ and 
	$1-z$, respectively. The left axis is the anomalous correction to the sixth-order 
	structure function $\Delta_6 = \zeta_6 - 2\zeta_3$ for the 
	forced-viscous system solution. The right axis is the scaling exponent $\alpha$ of 
	$|u_n^\pm| \sim k_n^\alpha$ for the instantonic solution. The gray line represents the Kolmogorov scaling $E_n \sim k_n^{-2/3}$.}
	\label{fig:7} 
	\end{figure}

The full Navier-Stokes dynamics corresponds to a mixture of the four 
helical classes, including models with all possible non-local interactions \cite{depietro2015inverse}. The behavior of the full coupled system may (or may not) 
inherit 
some properties of the individual models. In the shell model framework, this 
aspect can be studied by considering a linear combination of different models, 
e.g., by linearly coupling with a parameter $0 \le z \le 1$ the dynamical 
evolution of two models.
In Fig.~\ref{fig:7} we show the correlation between the anomalous correction to 
the sixth-order structure function $\Delta_6 = \zeta_6 - 2\zeta_3$ in the 
forced-viscous system and the scaling exponent $\alpha = 2(y-1)$ of $E_n$ measured in the instantonic solution, 
for a linear combination of model SM3 (no inverse cascade and regular 
instanton with a slope less steep than Kolmogorov's) with model SM2E (inverse cascade and chaotic instanton with a slope steeper than Kolmogorov's). The transition from the very intense (regular, small $|\alpha|$) to the weak (chaotic, large $|\alpha|$)  instanton and the transition from an intermittent to a nonintermittent dynamics in the forced-viscous 
regime occur at roughly the same value of $z$. The transition on the instanton slope is sharper. Moreover, the instanton becomes chaotic already for $z<0.1$, weakening the statement about the existence of a strict correlation among the presence of anomalous scaling and the  inviscid structure of the instantonic solutions observed  for the pure models ($z=0$ or 
$z=1$). Figure \ref{fig:7} also shows that the transition to an intermittent scaling ($-\Delta_6>0$) is observed for values of $\alpha$ around the Kolmogorov scaling ($-2/3$). Notice that there exists a residual intermittency even in the region where the instanton has a slope $|\alpha| > 2/3$ ($z \gtrsim 0.6 $). We cannot state if this effect is vanishing with increasing Reynolds number  because of numerical limitations.

\subsection{Energy transfer by Instantons}
In order to understand the transfer properties of instantonic solutions for 
each of the five models, we divided the shells inside the system into three 
domains:  the interval of shells $n$ where the instanton is initialized $I_0 = \{n_1 \le n \le n_2\}$ and 
 the interval of shells at larger and smaller scales, respectively, $I^< = \{n < n_1\}$ and 
$I^> = \{n > n_2\}$. For each instanton, we measured the energy contained in each of the three ranges at a late time $t^* \approx t_c$ (very large $\tau$). Normalizing this number by the total energy gives the 
fraction of energy transferred to larger and smaller scales, or kept in place, by a single instanton. Formally,
\begin{align}
\label{eq:transfer_up}
T_E^<(t^*) = \frac{1}{E} \sum_{n \in I^<} E_n(t^*) \, , \\
\label{eq:transfer_zero}
T_E^0 (t^*) = \frac{1}{E} \sum_{n \in I^0} E_n (t^*) \, , \\
\label{eq:transfer_down}
T_E^> (t^*) = \frac{1}{E} \sum_{n \in I^>} E_n(t^*) \, .
\end{align}

\begin{figure}
\includegraphics[width=0.75\linewidth]{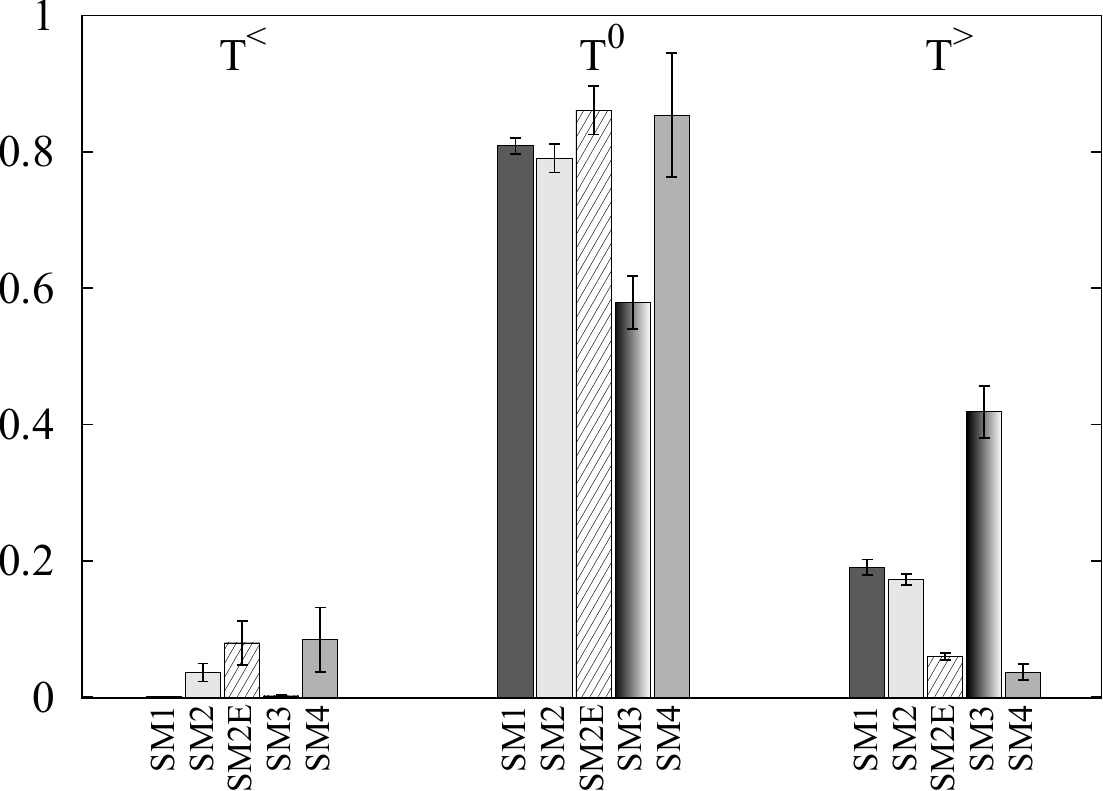}
\caption{Energy transfers (\ref{eq:transfer_up})--(\ref{eq:transfer_down}), 
from the scales where the initial condition is localized, towards larger 
$T^<$, smaller $T^>$, or the same scales $T^0$, for various models. The 
values are obtained by averaging over all the instantons in the ensemble and 
over different initial conditions (both differing in phases and in the number 
of shells where energy is initially present); the error bars show the 
standard deviation.}
\label{fig:8} 
\end{figure}

The transfers $T_E^{<}$, $T_E^{0}$, and $T_E^{>}$ for all the models, are 
shown in Fig. \ref{fig:8}. 
These transfers are averaged over all the instantons in the ensemble and over 
different choices for the 
width of the interval $I^0$ (ranging from three to seven shells). We see that in 
general instantons do not transfer forward (to small scales) a large amount of energy,
in agreement also with what is shown in Fig. 3, except for instantons of 
SM3 which are able to downscale almost $40\%$ of 
the initial energy.  For the backward energy transfer, the 
main effect is detected for models SM2E and SM4. Despite the fact 
that the total amounts of transferred energy are not too big, it is 
remarkable that the energy transfer by instantons follows the same direction as the energy cascade 
in stationary turbulent dynamics (see Table~\ref{tab:summary_dynamics}). In particular,
model SM3 has a very clear predominance in 
transferring forward. This might be considered a good indication that 
instantons play a relevant role in such helical interactions. For the inverse 
cascade, the properties of models SM2E and SM4 are less compelling but still present. 

\begin{table*}
	\caption{Summary of the dynamical properties of helical shell model 
		(\ref{eq:sabra_helical_general_up}) and (\ref{eq:sabra_helical_general_um})
		 and model 
		(\ref{eq:sabra_helical_2p2_up})and (\ref{eq:sabra_helical_2p2_um}) in 
		both the 
		blowup and stationary regimes. Here $\alpha$ is the scaling exponent of the energy spectrum for the instanton. We recall that in the stationary regime, a forward energy cascade induces a Kolmogorov spectrum $E_n \sim k_n^{-2/3}$.}
	\label{tab:summary_dynamics}
	\begin{tabular*}{\linewidth}{@{\extracolsep{\fill} } c | c c c | c c }
		\toprule
		& \multicolumn{3}{ c |}{Instanton dynamics} & \multicolumn{2}{ c 
		}{Stationary dynamics} \\
		Model &  Type  & Energy transfer & $\alpha$ & Intermittency & Energy dynamics \\ 
		\colrule
		SM1 & regular & borward & $-0.56$ & yes & forward cascade \\
		SM2 & chaotic & forward & $-1.3 < \alpha < -0.85$ & no & forward cascade \\
		SM2E & chaotic & backward & $-1.6 < \alpha < -1.1$ & no & backward cascade \\
		SM3 & regular & forward & $-0.03$ & yes & forward cascade \\
		SM4 & chaotic & backward & $\simeq -1.6$ & no & backward flux + 
		quasiequilibrium \\
		\botrule
	\end{tabular*}
\end{table*}

\begin{figure}
	\includegraphics[width=0.75\linewidth]{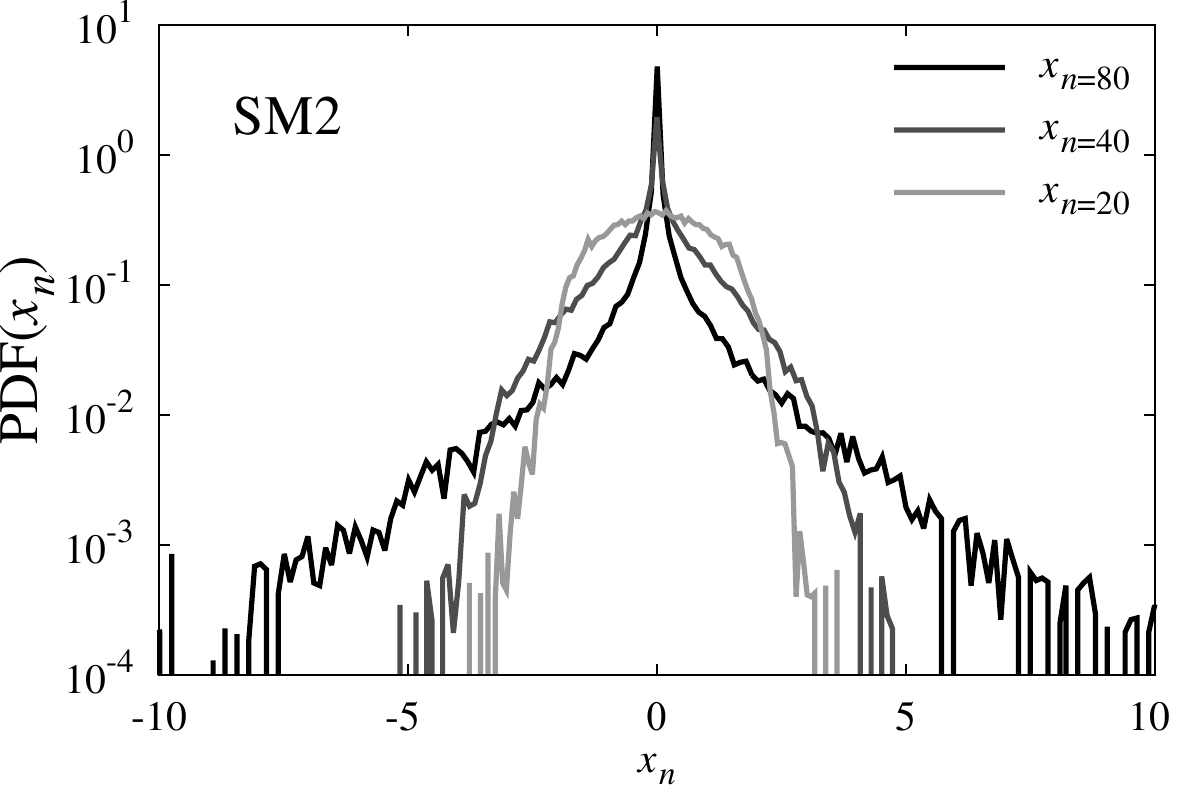} \\
	\includegraphics[width=0.75\linewidth]{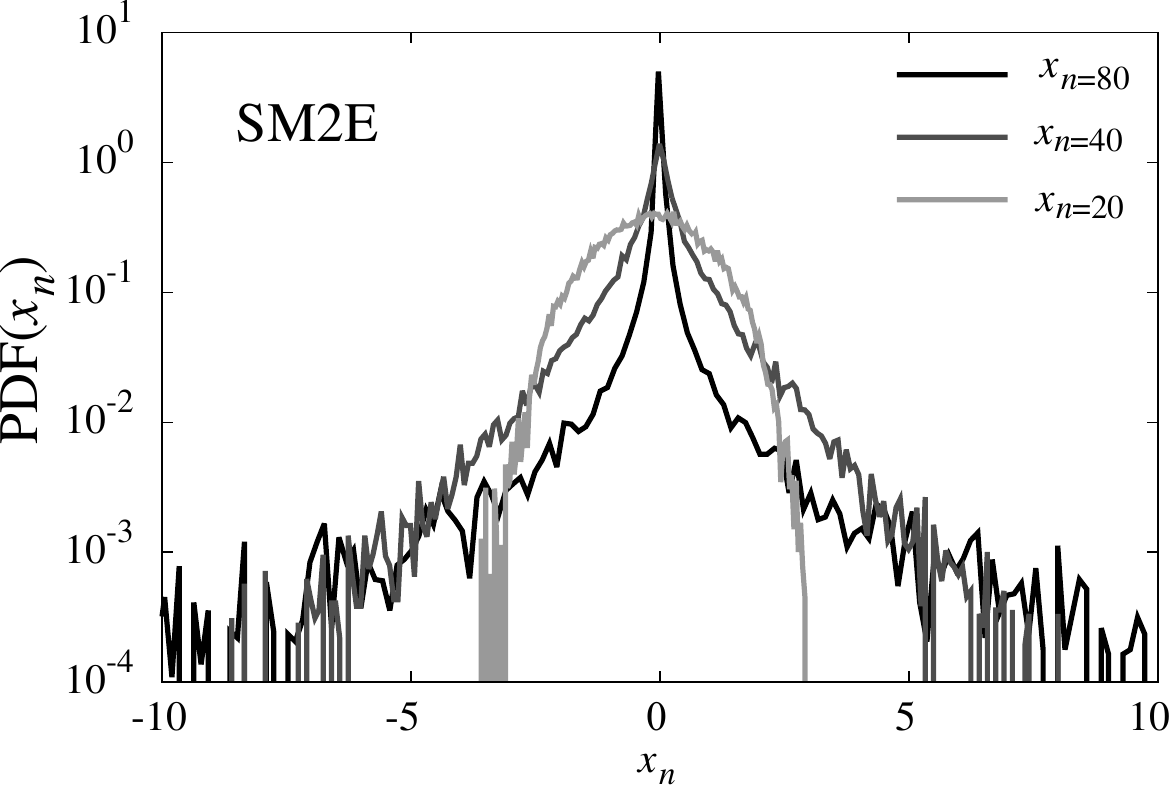} \\
	\includegraphics[width=0.75\linewidth]{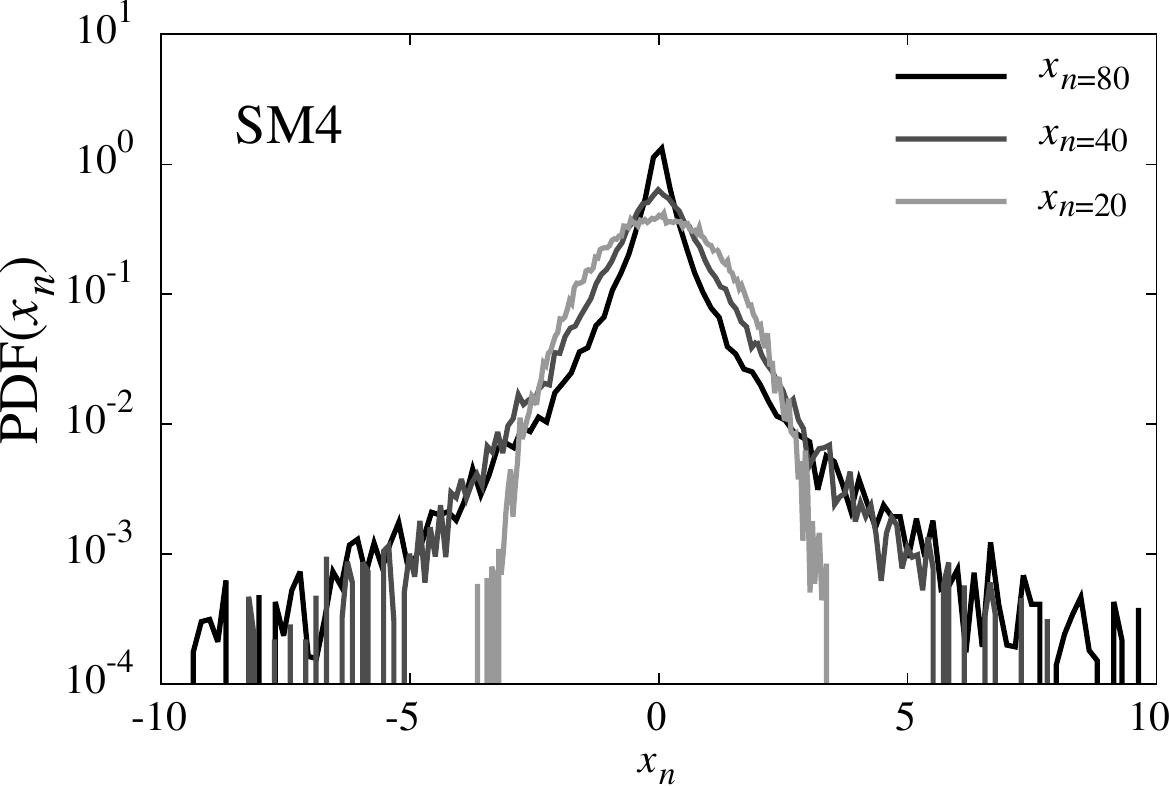} \\
	\caption{ PDF of $x_n \equiv \frac{\text{Re}[u_n] }{\sqrt{(\langle \text{Re}[u_n]^2 \rangle)}}$, ($\text{Re}$ is the real part), at different scales identified by the shell number 
	$n$, for models SM2, SM2E, and SM4.}
	\label{fig:9} 
\end{figure} 

\subsection{Intermittency of Instantons}
The presence of chaotic instantons  for models SM2, 
SM2E, and SM4 might eventually lead to nontrivial anomalous 
scaling by themselves, without considering the whole 
forced and viscous dynamics. In Fig. \ref{fig:9} we show the PDF of the real part of $u_n^+$ for different values of the shellnumber $n$, where the statistics is obtained over $O(10^6)$ instantons. As one can see, model SM2 and SM2E show a imperfect rescaling of the standardized PDF, even though the statistics does not allow one to make a firm statement about a strong breaking of self-similarity. The possibility that by allowing the instantons to travel for a much larger number of shells they all converge on one single averaged scaling exponent cannot be ruled out. This would indicate the existence of a chaotic attractor in the renormalized dynamics.

\section{Conclusions}
\label{sec:conclusions}
 We have studied the finite-time blowup solutions (instantons) for a 
set of four families of helical 
shell models that follow the exact decomposition 
of the Navier-Stokes equations in helical states. Four models SM1--SM4 were studied with the simplest short-range interactions allowed by the symmetries of the equations; an extra model SM2E was also considered with more nonlocal interactions  in order to study also  systems with an inverse energy cascade. When the models are initialized with energy at the large scales, 
the blowup solutions generate coherent structures that travel toward small 
scales. For models SM1 and SM3, the instantons are regular, less steep than the Kolmogorov scaling, and develop a self-similar asymptotic profile. For model SM4, the self-similarity holds only on average, with the instanton showing a chaotic evolution around a well defined mean profile. For models SM2 and SM2E, the instantons are again chaotic and oscillating among different states, apparently breaking a self-similar propagation, even on average. All models SM2, SM2E and SM4 have instantons with spectral slopes steeper than Kolmogorov.

The regularity or chaoticity of the blowup solutions correlates with the 
distinction of the various helical interactions based on the linear stability 
analysis of a single triad 
\cite{waleffe_1992_The_nature_of_triad_interactions}.
 In fact, the models with regular instantons (SM1 and SM3) belong to categories 
of helical interactions where the smallest wave number in a triad transfers 
energy to the other two, while the models with chaotic instantons (SM2, SM2E, and 
SM4) belong to categories in which the middle wave number in a triad 
transfers energy to the other two.

Another interesting correlation was observed concerning the intermittency in the 
stationary dynamics for the same helical shell models. In fact, the 
stationary regimes of both models SM1 and SM3 show 
anomalous scaling exponents for the velocity structure functions, 
quantitatively very similar to those of the Navier-Stokes 
turbulence~\cite{Lvov_1998_improved_shellmodels,benzi_1996_Helical_shell_models,
 biferale2003shell}. On the contrary, models SM2, SM2E, and 
SM4 do not show significant anomalous correction 
\cite{ditlevsen1996cascades, depietro2015inverse}. Combining two models, e.g., SM2E and SM3, one observes that whenever the small-scale stationary statistics is significantly intermittent, the instanton is less steep than the dimensional Kolmogorov scaling, independently of whether or not it is chaotic. This observation supports the idea that intermittency in the forced--viscous dynamics is influenced by instantons, if they are intense enough.

We also found a correlation between the energy transfers observed in the 
instantons and the energy fluxes measured in the stationary dynamics. All the 
models characterized by constant fluxes of energy toward small scales have 
instantons in which the dominant energy transfer is toward small 
scales and vice versa. 

Finally,  we have shown that 
model SM3 has a faster recovery of parity invariance at small scales compared to the 
other models. This seems to be the case also in the stationary dynamics 
\cite{rathmann2016role}. Furthermore, model SM3 is known to have 
a dynamics that is very robust with respect to variations in the model parameters such as 
the shell-to-shell ratio $\lambda$ or the dimensionality of the second 
inviscid invariant \cite{benzi_1996_Helical_shell_models}. All these clues 
reinforce the idea that the helical interaction present in model 
SM3 is actually the dominant component of the 3D Navier-Stokes 
dynamics.

\section*{Acknowledgments}
The authors acknowledge funding from the European Research Council under the 
European Union's Seventh Framework Programme, ERC Grant Agreement No 339032. A.A.M was supported by the CNPq Grant No. 302351/2015-9 and the FAPERJ Pensa Rio Grant No. E-26/210.874/2014.

\bibliography{bibliography}

\end{document}